\documentclass[11pt]{article}
\usepackage{DGfest,epsfig}
\usepackage{times}

\def\be{\begin{equation}}
\def\ee{\end{equation}}
\def\bea{\begin{eqnarray}}
\def\eea{\end{eqnarray}}

\begin{document}

\title{NONLOCAL CONDENSATES AND CURRENT-CURRENT CORRELATORS WITHIN THE INSTANTON LIQUID MODEL}
\author{ A.E. DOROKHOV }
\maketitle

\baselineskip 11.5pt

\address{Joint Institute for Nuclear Research, Bogoliubov Laboratory of Theoretical
Physics,\\
Dubna 141980, Russia}

\abstracts{The quark and gluon nonlocal condensates and current-current correlators
are discussed within the instanton liquid model.}

\section{Gluon field strength correlator.}

The non-perturbative vacuum of QCD is densely populated by long - wave
fluctuations of gluon and quark fields. The order parameters of this
complicated state are characterized by the vacuum matrix elements of various
singlet combinations of quark and gluon fields, condensates: $\left\langle :
\bar{q}q:\right\rangle $, ~$\left\langle :~F_{\mu \nu }^{a}F_{\mu \nu
}^{a}:\right\rangle $,$\left\langle :\bar{q}(\sigma _{\mu \nu }F_{\mu \nu
}^{a}\frac{\lambda ^{a}}{2})q:\right\rangle $ , {\it etc}. The nonzero quark
condensate $\left\langle :\bar{q}q:\right\rangle $ is responsible for the
spontaneous breakdown of chiral symmetry, and its value was estimated a long
time ago within the current algebra approach. The nonzero gluon condensate
$\left\langle :~F_{\mu \nu }^{a}F_{\mu \nu }^{a}:\right\rangle $ through
trace anomaly provides the mass scale for hadrons. Its value was
estimated within the QCD sum rule (SR) approach \cite{SVZ79} and first
evidence on its existence has been obtained in \cite{DiGiRos}. The values of low -
dimensional condensates were obtained phenomenologically from the QCD
SR analysis of the current-current correlators in various hadron channels.
In \cite{1,1a} it was proposed to study the effects of the gluon condensate on a
quarkonium state in order to estimate its absolute value.

The nonlocal vacuum condensates or vacuum correlators
\cite{NLC81,MihRad92} describe the distribution of quarks
and gluons in the non-perturbative vacuum. Physically, it means that vacuum
quarks and gluons can flow through the vacuum with nonzero momentum. From
this point of view the standard vacuum expectation values (VEVs) like
$\left\langle :\bar{q}q:\right\rangle $, $\left\langle :\bar{q}
D^{2}q:\right\rangle $~, $\left\langle :g^{2}F^{2}:\right\rangle ,\ \ldots $
appear as expansion coefficients of the quark $M(x)=\ \left\langle :~\bar{q}
(0)\hat{E}(0,x)q(x):\right\rangle $ and gluon $D^{\mu \nu ,\rho \sigma }(x) $
correlators in a Taylor series in the variable $x^{2}/4$.
The correlator $D^{\mu \nu ,\rho \sigma }(x)$ of gluonic field strengths
\begin{equation}
D^{\mu \nu ,\rho \sigma }(x-y)\equiv \left\langle :TrF^{\mu \nu }(x)\hat{E}
(x,y)F^{\rho \sigma }(y)\hat{E}(y,x):\right\rangle ,  \label{GluCor}
\end{equation}
may be parameterized in the form consistent with general requirements of the
gauge and Lorentz symmetries as
\cite{DoSi88}:
\begin{eqnarray}
D^{\mu \nu ,\rho \sigma }(x) &\equiv &\frac{1}{24}\left\langle
:F^{2}:\right\rangle \{(\delta _{\mu \rho }\delta _{\nu \sigma }-\delta
_{\mu \sigma }\delta _{\nu \rho })[D(x^{2})+D_{1}(x^{2})]+  \label{Fld_Cor}
\\
&+&(x_{\mu }x_{\rho }\delta _{\nu \sigma }-x_{\mu }x_{\sigma }\delta _{\nu
\rho }+x_{\nu }x_{\sigma }\delta _{\mu \rho }-x_{\nu }x_{\rho }\delta _{\mu
\sigma })\frac{\partial D_{1}(x^{2})}{\partial x^{2}}\},  \nonumber
\end{eqnarray}
where $\hat{E}(x,y)=P\exp \left( i\int_{x}^{y}A_{\mu }(z)dz^{\mu }\right) $
is the path-ordered Schwinger phase factor (the integration is performed
along the {\it straight} line) required for gauge invariance and $
\displaystyle A_{\mu }(z)=A_{\mu }^{a}(z)\frac{\lambda ^{a}}{2}$, ~$
\displaystyle F_{\mu \nu }(x)=F_{\mu \nu }^{a}(x)\frac{\lambda ^{a}}{2},$ $
~F_{\mu \nu }^{a}(x)=\partial _{\mu }A_{\nu }^{a}(x)-\partial _{\nu }A_{\mu
}^{a}(x)+f^{abc}A_{\mu }^{b}(x)A_{\nu }^{c}(x)$. The $P-$exponential ensures
the parallel transport of color from one point to other.
In (\ref{Fld_Cor}),
$\left\langle :F^{2}:\right\rangle =\left\langle :F_{\mu \nu }^{a}(0)F_{\mu
\nu }^{a}(0):\right\rangle $ is a gluon condensate, and $D(x^{2})$ and $
D_{1}(x^{2})$ are invariant functions which characterize nonlocal properties
of the condensate in different projections. The form factors are normalized
at zero by the conditions $D(0)=\kappa $, $D_{1}(0)=1-\kappa $, that depend
on the dynamics considered. For example, for the self-dual fields $\kappa =1$,
while in the Abelian theory without monopoles the Bianchi identity provides
$\kappa =0$.

In \cite{DEM97,DoLaur98}, one has shown that the instanton model of
the QCD vacuum provides a way to construct nonlocal vacuum condensates.
Within the effective single instanton (SI) approximation one has obtained
the expressions for the nonlocal gluon $D_{I}^{\mu \nu ,\rho \sigma }(x)$
and quark $M_{I}(x)$ condensates and derived the average virtualities of
quarks $\lambda _{q}^{2}$ and gluons $\lambda _{g}^{2}$ in the QCD vacuum.
The behavior of the correlation functions demonstrates that in the SI
approximation the model of nonlocal condensates can well reproduce the
behavior of the quark and gluon correlators at {\em short distances}.

In \cite{DEMM99}, it was suggested that the instanton
$A_{\mu }^{CI}(x)$ is developed in the physical vacuum field $b_{\mu }(x)$
interpolating large-scale vacuum fluctuations.
One has found that at small distances the instanton field dominates, and at
large distances it decreases exponentially $\sim
\exp \left[ -\frac{2}{3}\left( \eta _{g}\left|
x\right| \right) ^{3/2}\right] $ unlike the power-like decreasing SI. It is
important to note that the form of this asymptotics is independent of
the model for the background field and the driven parameter $\displaystyle
\eta _{g}\sim \left( \frac{N_{c}}{9\left( N_{c}^{2}-1\right) }R
\left\langle F_{b}^{2}\right\rangle _{b}\right) ^{\frac{1}{3}}$,
where $R$ is the correlation length and
$\left\langle F_{b}^{2}\right\rangle _{b}$ is the background field contribution to the
gluon condensate, only weakly depends on
it. This solution is called the constrained instanton (CI) \cite{Affl}.

The knowledge of the constraint-independent parts of CI allowed us to
construct the solution in the ansatz form
\begin{equation}
A_{\mu }^{CI,a}(x)=\overline{\eta }_{\nu \mu }^{a}\frac{x_{\nu }}{x^{2}}
\varphi _{g}\left( x^{2}\right),\ \ \ \ \ \  \
\varphi_g\left( x^{2}\right) =\frac{\overline{\rho }^{2}\left(
x^{2}\right) }{ x^{2}+\overline{\rho }^{2}\left( x^{2}\right)}
\label{CIanz}
\end{equation}
where the notation
$
\bar{\rho }^{2}\left( x^{2}\right) =a_{4/3}\eta _{g}^{2}x^{2}
K_{4/3}\left[\frac{2}{3}(\eta _{g}x)^{3/2} \right],
\bar{\rho }^{2}\left( 0\right) =\rho ^{2}
$
is introduced.
By translational invariance the center of CI can be shifted in
(\ref{CIanz}) from the origin to an arbitrary position $x_{0} $: $
x\rightarrow x-x_{0}$. The constrained instanton model introduces two characteristic scales
(correlation lengths). The short distance behavior of the
correlation functions is proportional to the instanton size and dominated by the single
instanton contribution.
The large correlation length is physically related to the confinement size $R$.
One sees that the $D\left( x^{2}\right)$ structure is close to the SI induced
function with the exponential asymptotics being developed at large distances.

\begin{figure}[h]
\hspace*{-1cm} \begin{minipage}{7cm}
\vspace*{0.5cm} \epsfxsize=6cm \epsfysize=5cm \centerline{\epsfbox{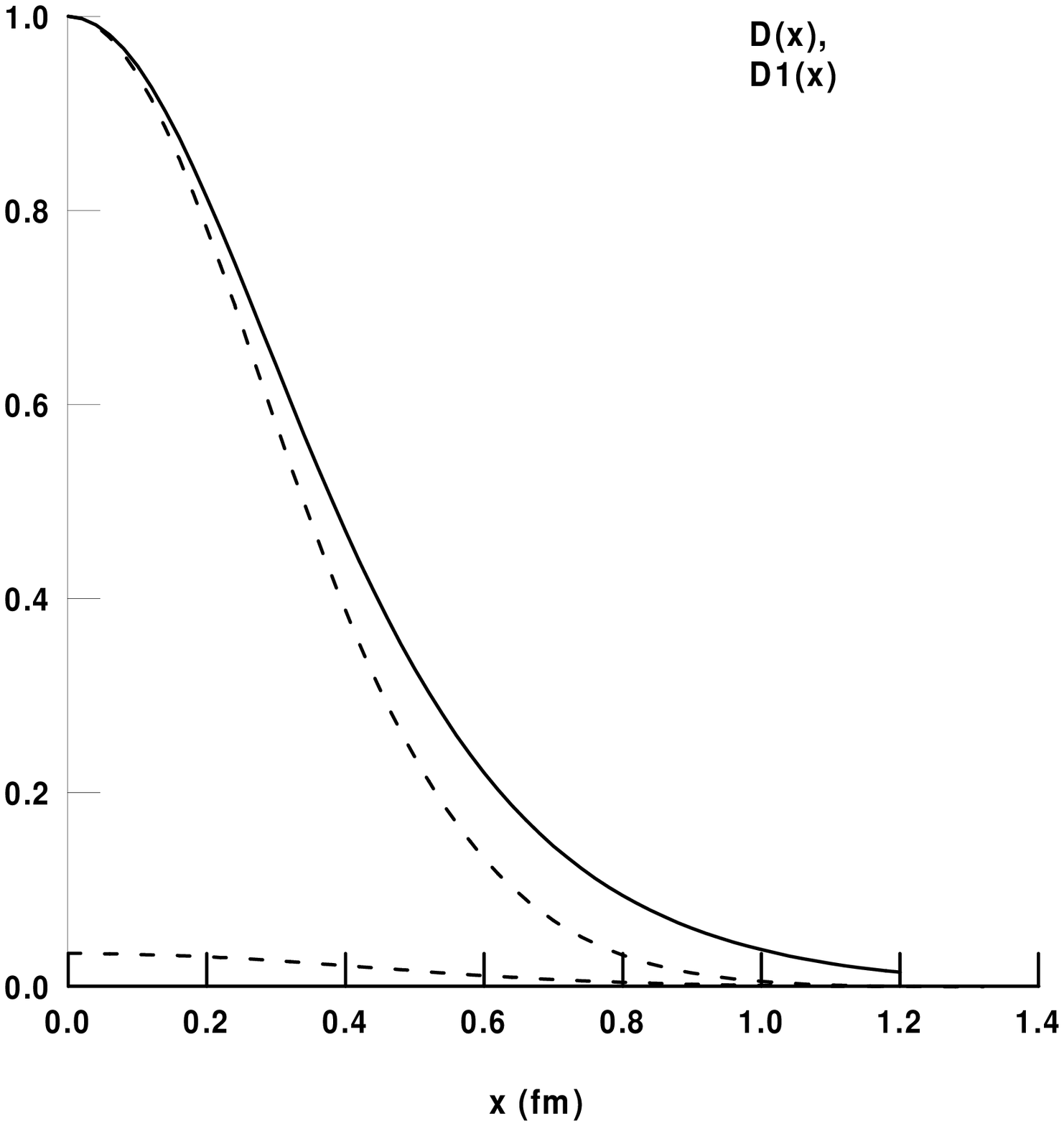}}
\caption[dummy0]{ The amplitudes $D$ (top lines) and $D_{1}$ (bottom
lines) (all normalized by $D(0)$) versus physical distance $x$, for the
instanton size $\protect\rho=0.3$ fm and parameters $(\protect\rho \protect
\eta _{g})^{2}=0$ (solid lines) and $(\protect\rho\protect\eta _{g})^{2}=1$
(dashed lines).
\label{Fig5} }
\end{minipage}\hspace*{0.5cm} \begin{minipage}{7cm}
\vspace*{0.5cm} \epsfxsize=6.5cm \epsfysize=6cm \centerline{\epsfbox
{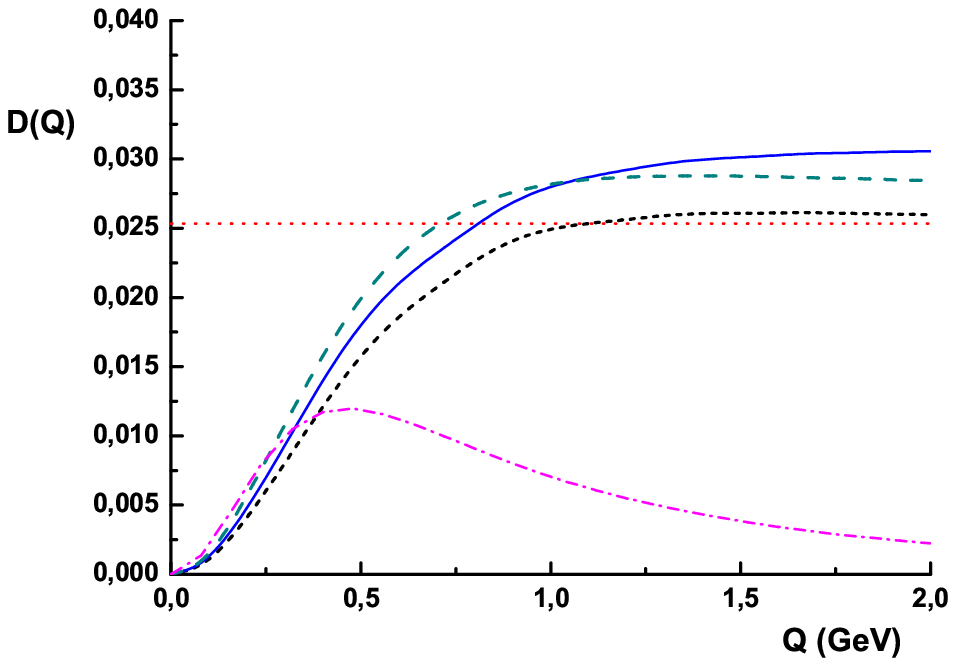}}
\caption[dummy0]{ The Adler function from
the ILM contributions: dynamical quark loop (short dashed), quark + chiral
loops + vector mesons (full line) versus the ALEPH data (dashed). The
dash-dotted line is the prediction of the constituent quark model (extended NJL) and
the dotted line is the asymptotic freedom prediction, $1/4\pi^{2}$.
\label{AdlVfig} }
\end{minipage}\end{figure}

 The gluon field strength correlation functions
$D\left( x^{2}\right) $ and $D_{1}\left(x^{2}\right) $  have been found
in \cite{DEMM99} and the results are shown in Fig. \ref{Fig5}.
We see that the results of the constrained instanton
model calculations are in a remarkable agreement with the results of lattice QCD simulations:
1) The $D_{1}\left( x^{2}\right) $ structure is much smaller than the $D\left( x^{2}\right) $
function \cite{DiGi97,DEMM99}; 2) the correlation length of the gluon field correlator is much smaller
than for the quark field correlator \cite{DiGi99,DEM97}; 3) the path dependence of the gluon field
strength correlator has similar features in both approach \cite{DiGi02,DoLaur04}.

Within the constrained instanton model there is a firm prediction \cite{DEMM99}, that
if the localized gluon classical field (constrained instanton) of the general form
\begin{equation}
A_\mu^{CI,a}(x)=\eta^a_{\nu\mu} \frac{x_\nu}{x^2}\varphi_g(x^2),
\end{equation}
with properties $\varphi_g(x^2\to 0)=1$ and $\varphi_g(x^2\to \infty) \to 0$
faster than any power of $1/x$,
then the correlation functions must have zeros in preasymptotic region
on the background of their exponential decay.
Thus the predicted asymptotics is of kind
\begin{equation}
D(x) \sim \exp{(-\Lambda x)} \cos{(b x+\delta(x))}
\end{equation}
rather than pure exponential one.
The appearance of oscillations is possible exclusively due to the Schwinger string phase
factor (link) and so is very interesting\footnote{In many works devoted to theoretical analysis of the
gluon field strength correlators the Schwinger link is neglected and so the important effect
of oscillations is missed.}.
 For interpretation of
these oscillations as specific "confining" behaviour of the propagator-like functions
see \cite{Maris95}.

The instanton model predicts the behaviour of nonperturbative part of gluon
correlation functions in the short and intermediate region assuming that
it is dominated by instanton vacuum component,
while the large-scale asymptotics is dominated by the background field.

The gluon correlators are the base elements of the stochastic
model of vacuum \cite{DoSi88}. They are also often used in the description of high-energy hadron
diffractive scattering amplitudes \cite{LN87,Nacht91}. However, as it was shown in \cite{DCh04} the
reduction of these amplitudes to the gluon
correlators is only valid for the Abelian model like that considered in \cite{LN87}.
As explicit calculations made within
the instanton model show in general case these amplitudes are process dependent \cite{DCh04,SH}.

\section{Hadronic current-current correlators.}

The transition from perturbative regime of QCD to nonperturbative one has yet
remained under discussion. At high momenta the fundamental degrees of freedom
are almost massless quarks. At low momenta the nonperturbative regime is
adequately described in terms of constituent quarks with masses dynamically
generated by spontaneous breaking of chiral symmetry. The instanton model of
QCD vacuum provides the mechanism of dynamical quark dressing in
the background of instanton vacuum and leads to generation of the momentum
dependent quark mass that interpolates these two extremes. Still it is not
clear how an intuitive picture of this transition may be tested at the level
of observables. Below we demonstrate that the Adler function and the amplitudes
related to the anomalous triangle diagram depending
on spacelike momenta may serve as the appropriate quantity.

The Adler function
defined as the logarithmic derivative of the current-current correlator can be
extracted from the experimental data of ALEPH and OPAL
 collaborations on inclusive hadronic $\tau$ decays. From
theoretical point of view it is well known that in high-energy asymptotically
free limit the Adler function calculated for massless quarks is a nonzero
constant. From the other side in the constituent quark model
this function is zero at zero virtuality. Thus the transition of
the Adler function from its constant asymptotic behaviour to zero is very
indicative concerning the nontrivial QCD dynamics at intermediate momenta. Below
we intend to show that the instanton liquid model which is a nonlocal chiral quark
model (ILM) describes this transition correctly \cite{ADprdG2}. The use in
the calculations of a covariant nonlocal low-energy quark model based on the
self-consistent approach to the dynamics of quarks has many attractive
features as it preserves the gauge invariance, is consistent with the
low-energy theorems, as well as takes into account the large-distance dynamics
controlled by the hadronic bound states.

In ILM in the chiral limit the
vector currents correlator has a transverse character \cite{DoBr03}
\begin{equation}
\Pi_{\mu\nu}^{v}\left(  Q^{2}\right)  =\left(  g_{\mu\nu}-\frac{q^{\mu}q^{\nu
}}{q^{2}}\right)  \Pi_{v}^{\mathrm{ILM}}\left(  Q^{2}\right)  , \label{PVmn}
\end{equation}
where the polarization function is given by the sum of the dynamical quark
loop, the intermediate vector mesons and the higher order mesonic
loop contributions.
The lowest order spectral representation of the polarization function consists
of zero width vector resonances  and two-meson states.
The dynamical quark loop
under condition of analytical confinement has no singularities in physical
space of momenta.

The dominant contribution to the vector current correlator at space-like
momentum transfer is given by the loop of light quarks with dynamical
momentum dependent mass $M(k)$
\cite{DoBr03} with the result\footnote{Within the context of ILM, the
integrals over the momentum are calculated by transforming the integration
variables into the Euclidean space, ($k^{0}\rightarrow ik_{4},$ $k^{2}
\rightarrow-k^{2}$).}
\begin{eqnarray}
\Pi_{V}^{Q\mathrm{Loop}}\left(  Q^{2}\right)   &&  =\frac{4N_{c}}{Q^{2}}
\int\frac{d^{4}k}{\left(  2\pi\right)  ^{4}}\frac{1}{D_{+}D_{-}}\left\{
M_{+}M_{-}+\left[  k_{+}k_{-}-\frac{2}{3}k_{\perp}^{2}\right]  _{ren}\right.
\label{Ploop}\\
& & +\left.  \frac{4}{3}k_{\perp}^{2}\left[  \left(  M^{\left(  1\right)
}\left(  k_{+},k_{-}\right)  \right)  ^{2}\left(  k_{+}k_{-}-M_{+}
M_{-}\right)  -\left(  M^{2}\left(  k_{+},k_{-}\right)  \right)  ^{\left(
1\right)  }\right]  \right\}  +\nonumber\\
& & +\frac{8N_{c}}{Q^{2}}\int\frac{d^{4}k}{\left(  2\pi\right)  ^{4}}
\frac{M\left(  k\right)  }{D\left(  k\right)  }\left[  M^{\prime}\left(
k\right)  -\frac{4}{3}k_{\perp}^{2}M^{\left(  2\right)  }\left(
k,k+Q,k\right)  \right]  ,\nonumber
\end{eqnarray}
where the notations
$k_{\pm}=k\pm Q/2,\qquad k_{\perp}^{2}=k_{+}k_{-}-\frac{\left(  k_{+}q\right)
\left(  k_{-}q\right)  }{q^{2}},
$
$
M_{\pm}=M(k_{\pm}),\ D(k)=k^2+M^2(k),D_{\pm}=D(k_{\pm})
$
are used. We also introduce the finite-difference derivatives defined for an
arbitrary function $F\left(  k\right)  $ as
\begin{equation}
F^{(1)}(k,k^{\prime})=\frac{F(k^{\prime})-F(k)}{k^{\prime2}-k^{2}},\qquad
F^{(2)}\left(  k,k^{\prime},k^{\prime\prime}\right)  =\frac{F^{(1)}
(k,k^{\prime\prime})-F^{(1)}(k,k^{\prime})}{k^{\prime\prime2}-k^{\prime2}}.
\label{FDD}
\end{equation}
The
expression for $\Pi_{V}^{Q\mathrm{Loop}}\left(  Q^{2}\right)  $ is formally
divergent and needs proper regularization and renormalization procedures which
are symbolically noted by $\left[  ..\right]  _{ren}$ for the divergent term.
At the same time the corresponding Adler function is well defined and finite.

Also we have checked that there is no pole in the vector correlator as
$Q^{2}\rightarrow0$, which simply means that photon remains massless with
inclusion of strong interaction. In the limiting cases the corresponding Adler function
satisfies general requirements of QCD
\begin{equation}
A_{V}^{\mathrm{ILM}}\left(  Q^{2}\rightarrow0\right)  =\mathcal{O}\left(
Q^{2}\right)  ,\qquad A_{V}^{\mathrm{ILM}}\left(  Q^{2}\rightarrow
\infty\right)  =\frac{N_{c}}{12\pi^{2}}+\frac{O_{2}^{V}}{Q^{2}}+\mathcal{O}
\left(  Q^{-4}\right)  .\label{Aasympt}
\end{equation}
The leading high $Q^{2}$ asymptotics comes from the $\left[  k_{+}k_{-}
-\frac{2}{3}k_{\perp}^{2}\right]  _{\mathrm{ren}}$ term in (\ref{Ploop}),
while the subleading asymptotics is driven by "tachionic" or $<A^2>$ term with
coefficient \cite{DoBr03}
\begin{equation}
O_{2}^{V}=-\frac{N_{c}}{2\pi^{2}}\int_{0}^{\infty}du\frac{uM\left(  u\right)
M^{\prime}\left(  u\right)  }{D\left(  u\right)  }.\label{Tachion}
\end{equation}
It is possible to integrate Eq. (\ref{Tachion}) in the dilute liquid approximation,
$u>>M^{2}(u)$,
\begin{equation}
O_{2}^{V}\approx\frac{N_{c}}{4\pi^{2}}M_{q}^{2}\approx4.7\cdot10^{-3}
\quad\mathrm{GeV}^{2},\label{O2}
\end{equation}
which is close to exact result \cite{DoBr03} and phenomenological estimate
from \cite{ChetNar}.

By using set of parameters found in ILM \cite{ADprdG2} the Adler function in the vector
channel is presented in Fig. \ref{AdlVfig}.

\section{$VA\widetilde{V}$ correlator}

Since discovery of anomalous properties \cite{Adler:1969gk,BJ} of the triangle
diagram with incoming two vector and one axial-vector currents
\cite{Rosenberg:1963pp} many new interesting results have been gained.
Recently the interest to triangle diagram has been renewed due to the problem
of accurate calculation of higher order hadronic contributions to muon
anomalous magnetic moment via the light-by-light scattering process,
that cannot be expressed as a
convolution of experimentally accessible observables and need to be estimated
from theory.

The triangle amplitude involving the axial current $A$ and two electromagnetic
currents (one soft $\widetilde{V}$ and one virtual $V$), which can be viewed as a mixing between
the axial and vector currents in the external electromagnetic field, were
considered recently in \cite{CMV,AD05 WLT,AD05wlts}. This amplitude can be written as
a correlator of the axial current $j_{\lambda}^{5}$ and two vector currents
$j_{\nu}$ and $\tilde{j}_{\mu}$
\begin{equation}
\widetilde{T}_{\mu\nu\lambda}=-\int\mathrm{d}^{4}x\mathrm{d}^{4}
y\,\mathrm{e}^{iqx-iky}\,\langle0|\,T\{j_{\nu}(x)\,\tilde{j}_{\mu
}(y)\,j_{\lambda}^{5}(0)\}|0\rangle\,,
\end{equation}
with the tilted current being
for the soft momentum photon vertex. In the specific kinematics when one
photon ($q_{2}\equiv q$) is virtual and another one ($q_{1}$) represents the
external electromagnetic field and can be regarded as a real photon with the
vanishingly small momentum $q_{1}$ depends only on two invariant functions,
longitudinal $w_{L}$ and transversal $w_{T}$ with respect to axial current
index,
\begin{eqnarray}
&&  \widetilde{T}_{\mu\nu\lambda}(q_{1},q)=\frac{1}{4\pi^{2}}\left[
-w_{L}\left(  q^{2}\right)  q^{\lambda}q_{1}^{\rho}q^{\sigma}\varepsilon
_{\rho\mu\sigma\nu}+\right. \nonumber\\
&&  \left.  +w_{T}\left(  q^{2}\right)  \left(  q^{2}q_{1}^{\rho}
\varepsilon_{\rho\mu\nu\lambda}-q^{\nu}q_{1}^{\rho}q_{2}^{\sigma}
\varepsilon_{\rho\mu\sigma\lambda}+q^{\lambda}q_{1}^{\rho}q_{2}^{\sigma
}\varepsilon_{\rho\mu\sigma\nu}\right)  \right]  . \label{Tt}
\end{eqnarray}
Both structures are transversal with respect to vector current, $q^{\nu
}\widetilde{T}_{\mu\nu\lambda}=0$. As for the axial current, the first
structure is transversal with respect to $q^{\lambda}$ while the second is
longitudinal and thus anomalous.

In the local theory the one-loop result for the invariant functions $w_{T}$
and $w_{L}$ is\footnote{Here and below the small effects of isospin violation
is neglected, considering $m_{f}\equiv m_{u}=m_{d}$.}
\begin{equation}
w_{L}^{\mathrm{1-loop}}=2\,w_{T}^{\mathrm{1-loop}}=\frac{2N_{c}}{3}\int
_{0}^{1}\frac{\mathrm{d}\alpha\,\alpha(1-\alpha)}{\alpha(1-\alpha)q^{2}
+m_{f}^{2}}\,, \label{wlt}
\end{equation}
where the factor $N_{c}/3$ is due to color number and electric charge. In the
chiral limit, $m_{f}=0$, one gets the result for space-like momenta $q$
$\left(  q^{2}\geq0\right)  $
$
w_{L}\left(  q^{2}\right)  =2w_{T}\left(  q^{2}\right)  ={2}/{q^{2}}.
\label{WLTch}
$

The appearance of the longitudinal structure is the consequence of the axial
Adler-Bell-Jackiw anomaly \cite{Adler:1969gk,BJ}. For the nonsinglet axial
current $A^{\left(  3\right)  }$ there are no perturbative \cite{Adler:er} and
nonperturbative \cite{tHooft} corrections to the axial anomaly and, as
consequence, the invariant function $w_{L}^{\left(  3\right)  }$ remains
intact when interaction with gluons is taken into account. Recently, it was
shown that the relation
\begin{equation}
w_{LT}\left(  q^{2}\right)  \equiv w_{L}\left(  q^{2}\right)  -2w_{T}\left(
q^{2}\right)  =0, \label{wtwl}
\end{equation}
which holds in the chiral limit at the one-loop level (\ref{WLTch}), gets no
perturbative corrections from gluon exchanges in the iso-triplet case
\cite{VainshPLB03}. Nonperturbative nonrenormalization of the nonsinglet
longitudinal part follows from the 't~Hooft consistency condition
\cite{tHooft}, i.e. the exact quark-hadron duality realized as a
correspondence between the infrared singularity of the quark triangle and the
massless pion pole in terms of hadrons. OPE\ analysis indicates that at large
$q$ the leading nonperturbative power corrections to $w_{T}$ can only appear
starting with terms $\sim1/q^{6}$ containing the matrix elements of the
operators of dimension six \cite{Knecht02}. Thus, the transversal part of the
triangle with a soft momentum in one of the vector currents has no
perturbative corrections nevertheless it is modified nonperturbatively.
However, for the singlet axial current $A^{\left(  0\right)  }$ due to the
gluonic $U_{A}\left(  1\right)  $ anomaly there is no massless state even in
the chiral limit. Instead, the massive $\eta^{\prime}$ meson appears. So, one
expects nonperturbative renormalization of the singlet anomalous amplitude
$w_{L}^{\left(  0\right)  }$ at momenta below $\eta^{\prime}$ mass. Below we
demonstrate how the anomalous structure $w_{L}^{\left(  3\right)  }$ is
saturated within the instanton liquid model. We also calculate the transversal
invariant function $w_{T}$ at arbitrary space-like $q$ and show that within
the instanton model in the chiral limit at large $q^{2}$ all allowed by OPE
power corrections to $w_{T}$ cancel each other and only exponentially
suppressed corrections remain \cite{AD05 WLT,AD05wlts}. The nonperturbative
corrections to $w_{T}$ at large $q^{2}$ have exponentially decreasing behavior
related to the short distance properties of the instanton nonlocality in the
QCD vacuum.

Within the instanton liquid model the nondiagonal ($VA\widetilde{V}$) correlator
of vector current and
nonsinglet axial-vector current in the external electromagnetic field
is given by
\begin{eqnarray}
& & \widetilde{T}_{\mu\nu\lambda}(q_{1},q_{2})=-2N_{c}\int\frac{d^{4}k}{\left(
2\pi\right)  ^{4}}Tr\left[  \Gamma_{\mu}\left(  k+q_{1},k\right)  S\left(
k+q_{1}\right)  \right.  \cdot\nonumber\\
& & \left.  \cdot\Gamma_{\lambda}^{5}\left(  k+q_{1},k-q_{2}\right)  S\left(
k-q_{2}\right)  \Gamma_{\nu}\left(  k,k-q_{2}\right)  S\left(  k\right)
\right]  , \label{Tncqm}
\end{eqnarray}
where the quark propagator $S^{-1}(k)=\hat k-M(k)$. The effective vector and axial-vector
vertices consist of the local and nonlocal parts and satisfy the Ward-Takahashi
identities \cite{DoBr03,ADoLT00}. The
structure of the vector vertices guarantees that the amplitude is transversal
with respect to vector indices
$
\widetilde{T}_{\mu\nu\lambda}(q_{1},q_{2})q_{1}^{\mu}=\widetilde{T}_{\mu
\nu\lambda}(q_{1},q_{2})q_{2}^{\nu}=0
$
and the Lorentz structure of the amplitude is given by (\ref{Tt}).

The contribution of the diagram where all vertices are local to the invariant functions at
space-like momentum transfer, $q^{2}\equiv q_{2}^{2}$, are given by
\begin{equation}
w_{L}^{(loc)}\left(  q^{2}\right)  =\frac{4N_{c}}{9q^{2}}\int\frac{d^{4}k}
{\pi^{2}}\frac{1}{D_{+}^{2}D_{-}}\left[  k^{2}-4\frac{\left(  kq\right)  ^{2}
}{q^{2}}+3\left(  kq\right)  \right]  , \label{A46a}
\end{equation}
\begin{equation}
w_{LT}^{\left(  loc\right)  }\left(  q^{2}\right)  =0, \label{A6a}
\end{equation}
where we also consider the combination of invariant functions $w_{LT}$, (\ref{wtwl}), which show up
nonperturbative dynamics.
The notations used here and below are
$
k_{+}=k, k_{-}=k-q, k_{\perp}^{2}=k_{+}k_{-}-\frac{\left(
k_{+}q\right)  \left(  k_{-}q\right)  }{q^{2}}.
$
At large $q^{2}$ one gets
$
w_{L}^{(loc)}\left(  q^{2}\rightarrow\infty\right)  =\frac{2N_{c}}{3}
\frac{1}{q^{2}}  \label{A4as}
$
showing saturation of the anomaly. The reason is that the leading large $q^{2}$ asymptotics
of (\ref{A46a}) is given
by the configuration where the large momentum is passing through all quark
lines. Then the dynamical quark mass $M(k)$ reduces to zero and the asymptotic
limit of triangle diagram with dynamical quarks and local vertices coincides
with the standard triangle amplitude with massless quarks and, thus, it is
independent of the model.

The contribution to the form factors when the nonlocal parts of the vector and
axial-vector vertices are taken into account is given by
\begin{eqnarray}
w_{L}^{\left(  nonloc\right)  }\left(  q^{2}\right)   &&  =\frac{4N_{c}}
{3q^{2}}\int\frac{d^{4}k}{\pi^{2}}\frac{1}{D_{+}^{2}D_{-}}\left\{
M_{+}\left[  M_{+}-\frac{4}{3}M_{+}^{\prime}k_{\perp}^{2}\right]  -\right.
\nonumber\\
&&  \left.  -M^{2(1)}(k_{+},k_{-})\left(  2\frac{\left(  kq\right)  ^{2}}
{q^{2}}-\left(  kq\right)  \right)  \right\}  . \label{A4bt}
\end{eqnarray}

\begin{figure}[h]
\hspace*{-1cm} \begin{minipage}{7cm}
\vspace*{0.5cm} \epsfxsize=6cm \epsfysize=5cm \centerline{\epsfbox{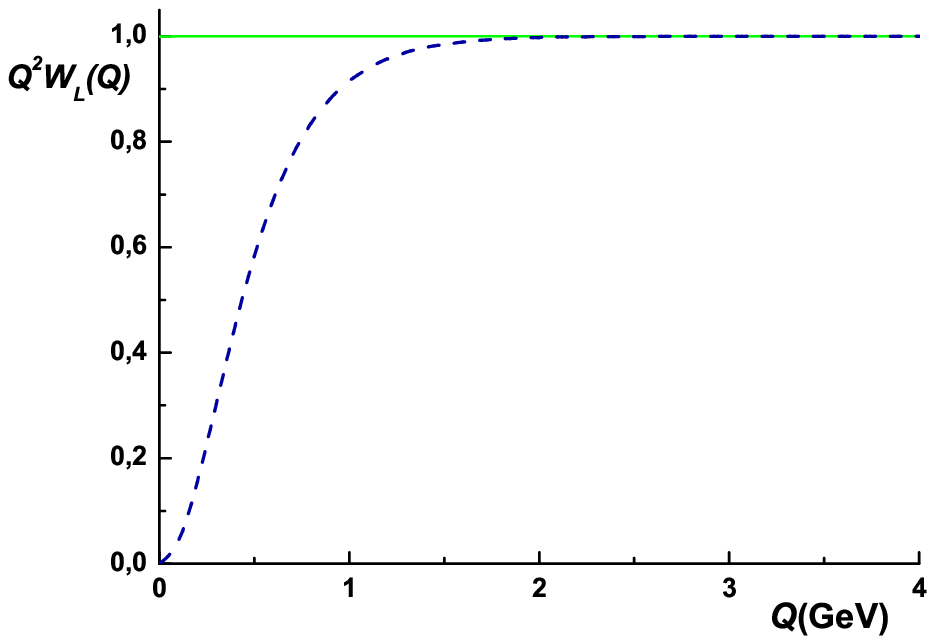}}
\caption[dummy0]{ Normalized $w_L$
invariant function in the nonsinglet case (solid line)
and singlet case (dashed line).
\label{WLfig} }
\end{minipage}\hspace*{0.5cm} \begin{minipage}{7cm}
\vspace*{0.5cm} \epsfxsize=6cm \epsfysize=5cm \centerline{\epsfbox
{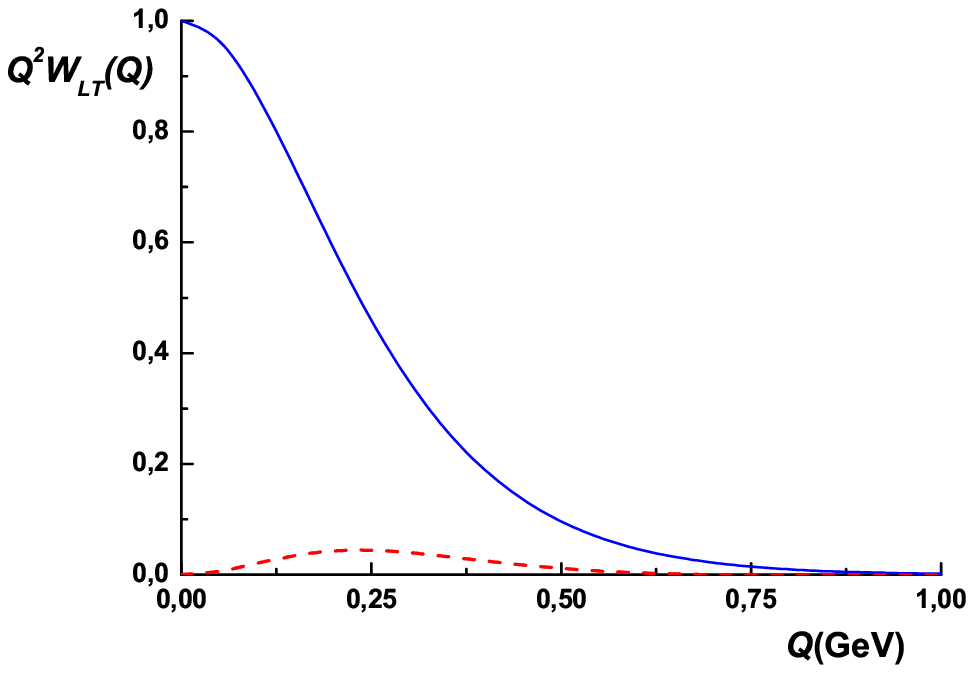}}
\caption[dummy0]{ Normalized $w_{LT}$
invariant function  versus
$Q$ predicted by the instanton model in the nonsinglet case (solid line)
and singlet case (dashed line).
\label{WLTfig} }
\end{minipage}\end{figure}

Summing analytically the local (\ref{A46a}) and nonlocal (\ref{A4bt}) parts
provides us with the result required by the axial anomaly \cite{AD05 WLT}
\begin{equation}
w_{L}(q^{2})=\frac{2N_{c}}{3}\frac{1}{q^{2}}. \label{A4Tot}
\end{equation}
Fig. \ref{WLfig} illustrates saturation of the anomaly in the non-singlet and singlet
cases.
Note, that at zero virtuality the saturation of anomaly follows from anomalous
diagram of pion decay in two photons. This part is due to the triangle diagram
involving nonlocal part of the axial vertex and local parts of the photon
vertices. The result (\ref{A4Tot}) is in agreement with the statement about
absence of nonperturbative corrections to longitudinal invariant function
following from the 't Hooft duality arguments.

For $w_{LT}(q^{2})$ a number of cancellations takes place and the final result
is quite simple \cite{AD05 WLT}
\begin{eqnarray}
&&  w_{LT}\left(  q^{2}\right)  =\frac{4N_{c}}{3q^{2}}\int\frac{d^{4}k}{\pi
^{2}}
\frac{\sqrt{M_{-}}}{D_{+}^{2}D_{-}}\left\{  \sqrt{M_{-}}\left[
M_{+}-\frac{2}{3}M_{+}^{\prime}\left(  k^{2}+2\frac{\left(  kq\right)  ^{2}
}{q^{2}}\right)  \right]  -\right. \nonumber\\
&&  \left.  -\frac{4}{3}k_{\perp}^{2}
\left[  \sqrt{M_{+}}M^{(1)}(k_{+},k_{-})-2\left(  kq\right)
M_{+}^{\prime}\sqrt{M}^{\left(  1\right)  }(k_{+},k_{-})\right]  \right\}  .
\label{wLT}
\end{eqnarray}

The behavior of $w_{LT}(q^{2})$ is presented in Fig. \ref{WLTfig}. In the
above expression the integrand is proportional to the product of nonlocal form
factors $f\left(  k_{+}^{2}\right)  f\left(  k_{-}^{2}\right)  $ depending on
quark momenta passing through different quark lines. Then, it becomes evident
that the large $q^{2}$ asymptotics of the integral is governed by the
asymptotics of the nonlocal form factor $f\left(  q^{2}\right)  $ which is
exponentially suppressed \cite{DEM97,AD05 WLT}. Thus, within the instanton model the
distinction between longitudinal and transversal parts is exponentially
suppressed at large $q^{2}$ and all allowed by OPE power corrections are
canceled each other. The instanton liquid model indicates that it may be
possible that due to the anomaly the relation (\ref{wtwl}) is violated at
large $q^{2}$ only exponentially.

The calculations of the singlet $VA\widetilde{V}$ correlator results in the
following modification of the nonsinglet amplitudes \cite{AD05wlts}
\begin{eqnarray}
w_{L}^{\left(  0\right)  }(q^{2})  &&  =\frac{5}{3}w_{L}^{\left(  3\right)
}\left(  q^{2}\right)  +\Delta w^{\left(  0\right)  }\left(  q^{2}\right)
,\label{WL0}\\
w_{LT}^{\left(  0\right)  }(q^{2})  &&  =\frac{5}{3}w_{LT}^{\left(  3\right)
}\left(  q^{2}\right)  +\Delta w^{\left(  0\right)  }\left(  q^{2}\right)  ,
\label{WLT0}
\end{eqnarray}
where
\begin{eqnarray}
\Delta w^{\left(  0\right)  }\left(  q^{2}\right)   & & =-\frac{5N_{c}}{9q^{2}
}\frac{1-G^{\prime}/G}{1-G^{\prime}J_{PP}\left(  q^{2}\right)  }\int
\frac{d^{4}k}{\pi^{4}}\frac{\sqrt{M_{+}M_{-}}}{D_{+}^{2}D_{-}}\left[
M_{+}-\frac{4}{3}M_{+}^{\prime}k_{\perp}^{2}-\right. \nonumber\\
& & \left.  -M^{(1)}(k_{+},k_{-})\left(  \frac{4}{3}\frac{\left(  kq\right)
^{2}}{q^{2}}+\frac{2}{3}k^{2}-\left(  kq\right)  \right)  \right]  .
\label{DW0}
\end{eqnarray}

Fig. \ref{WLfig} illustrates how the singlet longitudinal amplitude
$w_{L}^{\left(  0\right)  }$ is renormalized at low momenta by the presence of
the $U_{A}\left(  1\right)  $ anomaly. The behavior of $w_{LT}^{\left(
0\right)  }(q^{2})$ is presented in Fig. \ref{WLTfig}. Precise form and even
sign of $w_{LT}^{\left(  0\right)  }(q^{2})$ strongly depend on the ratio of
singlet and nonsinglet couplings $G^{\prime}/G$ and has to be defined in the calculations with more
realistic choice of model parameters.

\section{Conclusions}
We have analyzed the gluon field strength correlator (nonlocal gluon condensate),
the vector Adler function and the nondiagonal vector - axial-vector correlator in
external field (anomalous triangle amplitude) for Euclidean
(spacelike) momenta within an effective nonlocal chiral quark model motivated
by the instanton model of QCD vacuum. The results on the gluon correlator are in good qualitative
agreement with the lattice QCD simulations.
The dominant contributions to the Adler function and the triangle amplitude
come from the loop contributions of the light
quark with dynamical momentum dependent mass.
It is this contribution that provides the matching between
low energy hadronized phase and the high energy QCD which is clearly seen in
the behaviour of the Adler function (Fig. \ref{AdlVfig}) and the $w_{LT}$ combination of
the triangle amplitudes (Fig. \ref{WLTfig}).
The results obtained are close to estimates of the vector Adler
function extracted directly from the ALEPH data on
hadronic inclusive $\tau$ decays and transformed by dispersion relations to
the spacelike region.

\section*{Acknowledgments}
The author is grateful to A. Di Giacomo, N. I. Kochelev, S. V. Mikhailov
for helpful discussions on the subject of the present work.
The author also thanks for partial support from
the Russian Foundation for Basic Research projects nos. 03-02-17291,
04-02-16445.

\end{document}